\documentclass[sigconf]{acmart}
\AtBeginDocument{%
  \providecommand\BibTeX{{%
    \normalfont B\kern-0.5em{\scshape i\kern-0.25em b}\kern-0.8em\TeX}}}

\renewcommand\footnotetextcopyrightpermission[1]{}
\setcopyright{none}

\usepackage{url}
\usepackage{hyperref}
\usepackage{graphicx}%
\usepackage{multirow}%
\usepackage{amsthm}%
\usepackage{mathrsfs}%
\usepackage[title]{appendix}%
\usepackage{xcolor}%
\usepackage{textcomp}%
\usepackage{booktabs}%
\usepackage{algorithm}%
\usepackage{algorithmicx}%
\usepackage{algpseudocode}%
\usepackage{listings}%
\usepackage{xspace}
\usepackage{subcaption}
\usepackage{enumitem}
\usepackage{csquotes}
\usepackage{booktabs}
\usepackage{tikz}
\usepackage{xurl}

\usepackage{rotating}

\begin{document}

%% Cover page for the arXiv copy: reference to the original ACM
%% publication, ahead of the paper's own title page.
\thispagestyle{empty}
\begin{center}
\vspace*{3cm}
{\Large\bfseries A Context-Aware Cultural Heritage Guide Powered by LLMs}\\[1.5em]
Liliana Ardissono, Fabio Ferrero, Angelo Geninatti Cossatin, Claudio Mattutino, Noemi Mauro\\[3em]
\textit{This is the authors' version of the work.}\\[1.5em]
The definitive Version of Record was published in\\
Proceedings of the 34th ACM Conference on User Modeling, Adaptation and Personalization (UMAP~'26),\\
June 08--11, 2026, Gothenburg, Sweden.\\[1em]
DOI: \url{https://doi.org/10.1145/3774935.3812714}\\[3em]
\copyright{} Licensed under CC BY-NC-ND 4.0.
\end{center}
\newpage

\title{A Context-Aware Cultural Heritage Guide Powered by LLMs}

\author{Liliana Ardissono}
 \orcid{0000-0002-1339-4243}
 \affiliation{%
   \institution{Università degli Studi di Torino}
   \streetaddress{Corso Svizzera 185}
   \city{Torino} 
   \state{} 
   \country{Italy}
   \postcode{10149}
 }
 \email{liliana.ardissono@unito.it}

 \author{Fabio Ferrero}
 \orcid{0009-0007-2941-5030}
 \affiliation{%
   \institution{Università degli Studi di Torino}
   \streetaddress{Corso Svizzera 185}
   \city{Torino} 
   \state{} 
   \country{Italy}
   \postcode{10149}
 }
 \email{fab.ferrero@unito.it}

 \author{Angelo Geninatti Cossatin}
 \orcid{0009-0007-5378-7061}
 \affiliation{%
   \institution{Università degli Studi di Torino}
   \streetaddress{Corso Svizzera 185}
   \city{Torino} 
   \state{} 
   \country{Italy}
   \postcode{10149}
 }
\email{angelo.geninatticossatin@unito.it}

\author{Claudio Mattutino}
 \orcid{0000-0002-0413-2436}
 \affiliation{%
   \institution{Università degli Studi di Torino}
   \streetaddress{Corso Svizzera 185}
   \city{Torino} 
   \state{} 
   \country{Italy}
   \postcode{10149}
 }
 \email{claudio.mattutino@unito.it}

\author{Noemi Mauro}
 \orcid{0000-0001-8234-3266}
 \affiliation{%
   \institution{Università degli Studi di Torino}
   \streetaddress{Corso Svizzera 185}
   \city{Torino} 
   \state{} 
      \country{Italy}
   \postcode{10149}
 }
\email{noemi.mauro@unito.it}

\renewcommand{\shortauthors}{L. Ardissono et al.}
\renewcommand{\shorttitle}{A Context-Aware Cultural Heritage Guide Powered by LLMs}

%%
%%Workshop title and acronym, along with an abstract of the workshop (200 words maximum). Please note that this title and abstract will be used as your short workshop description on the website upon acceptance. For each organizer, include name, affiliation, and primary email address.
\begin{abstract}
We present an extension of Triangolazioni (a Cultural Heritage webapp) to enrich curated content with context-dependent, external information provided by Large Language Models (LLMs) within a loosely-coupled architecture agnostic to the LLM. 
The system supports context-dependent information search and presentation within an architecture agnostic to the exploited LLM.
\end{abstract}





\maketitle
\let\thefootnote\relax
\footnotetext{© 2026 Copyright held by the owner/author(s). Licensed under CC BY-NC-ND 4.0. Published at UMAP '26: Proceedings of the 34th ACM Conference on User Modeling, Adaptation and Personalization, DOI: https://doi.org/10.1145/3774935.3812714}

\section{Introduction}
Large Language Models (LLMs) open new possibilities for developing Cultural Heritage (CH) guides based on chatbots, thanks to their strength in understanding and answering users' questions, and their ability to gather information from heterogeneous data sources. However, they limit interaction with users to message-based conversations. Moreover, they have to be frequently replaced to keep pace with the quick evolution of Generative AI technology towards more reliable models. 

Early research on Cultural Heritage exploration investigated advanced information filtering and presentation techniques to enhance user experience when navigating curated physical or virtual CH sites \citep{Kuflik-etal:11,deCarolis-etal:18}.
%Other works enhanced the user experience through Augmented Reality or Virtual Reality \citep{Bonis-etal:09,Fenu-Pittarello:18,Bekele-etal:18}. 
Chatbots have been introduced in CH sites to simplify the interaction with the user, but they work on specific knowledge bases. 
%The chatbot is the guide's user interface and works as an ``info bot” to help users plan their visit \citep{Tzouganatou:18} by conversing with them.
For example, \cite{Machidon-etal:20} proposed a chatbot based on Google's DialogFlow to assist users in exploring the content of the Europeana ontology. 
%\cite{Casillo-etal:22} developed an ontology-guided chatbot that presents information about the Archaeological Urban Park of Naples.
%Furthermore, some research leveraged Virtual Reality to model chatbots as Virtual Humans \citep{Noh-Hong:21,Sylaiou-Fidas:22, Chalmers-etal:21} and Audio Augmented Reality to provide an immersive interaction environment \citep{Tsepapadakis-Gavalas:23}.
%Furthermore, researchers developed embodied chatbots for museums and focused on the impact of the interaction style and modality in learning environments \citep{Noh-Hong:21}. 
To expand the available information, external data sources have been integrated through Semantic Web technologies \citep{Faralli-etal:22,Kim-etal:17,Rinaldi-etal:22}. 
%Moreover, \cite{Varitimiadis-etal:21} proposed to exploit graph-based, distributed, and collaborative multi-chatbot conversational AI systems based on knowledge graphs. 
However, this integration still limits the knowledge bases to a closed set of resources. 

LLMs can acquire information from public repositories to extend their knowledge. However, they are rarely used in CH systems. Moreover, systems that leverage this technology (e.g., \cite{Trichopoulos-etal:23a}) raise two main concerns. 
First, as the LLM represents the user interface, it constrains the interaction with the user to the exchange of messages.
Second, the LLM is typically fine-tuned to the system's knowledge base to enhance its factual accuracy. Thus, the system is tied to it. 

In contrast, our work integrates web-based presentation and conversation with the LLM within a loosely-coupled architecture to facilitate LLM replacement. 

This paper presents a system that integrates a web-based cultural heritage site with an LLM-powered chatbot that extends curated content with external data, enriching the information presented beyond closed knowledge bases. By tracking the user's navigation of the website, the system supports content exploration through question formulation support and personalized question-answering. The system architecture, developed for \citep{Geninatti-etal:25}, provides a loosely-coupled integration of LLMs into a CH website. It synchronizes the chatbot with the user's browsing activity in the website through a shared context that makes the system agnostic to the LLM.

The novel features of our work are:
\begin{enumerate}
\item
A context-aware question-answering function based on LLMs to satisfy individual information needs beyond a CH website's curated content. 
\item
A context-dependent suggestion of questions to help the user expand the visited content with relevant information.
\item
A loosely coupled architectural model to flexibly integrate LLMs in a CH website.
\end{enumerate}
%In \citep{Geninatti-etal:25}, we found that this function enhances user experience, especially for people with low curiosity levels (according to Curiosity and Exploration Inventory-II - CEI-II) who are guided in formulating effective questions. 

%\textcolor{brown}{The described system is available at \url{https://exptriangolazioni.ontomap.eu/}. Before UMAP, the connection to the LLM will be frozen. Thus, the system will present static questions and answers.}
A video of the interaction with the system is available at the following link: \url{https://youtu.be/Vf_5yhndMGk}.

\begin{figure*}
     \centering
     \begin{subfigure}[b]{0.49\textwidth}
         \centering
         \includegraphics[width=\textwidth]{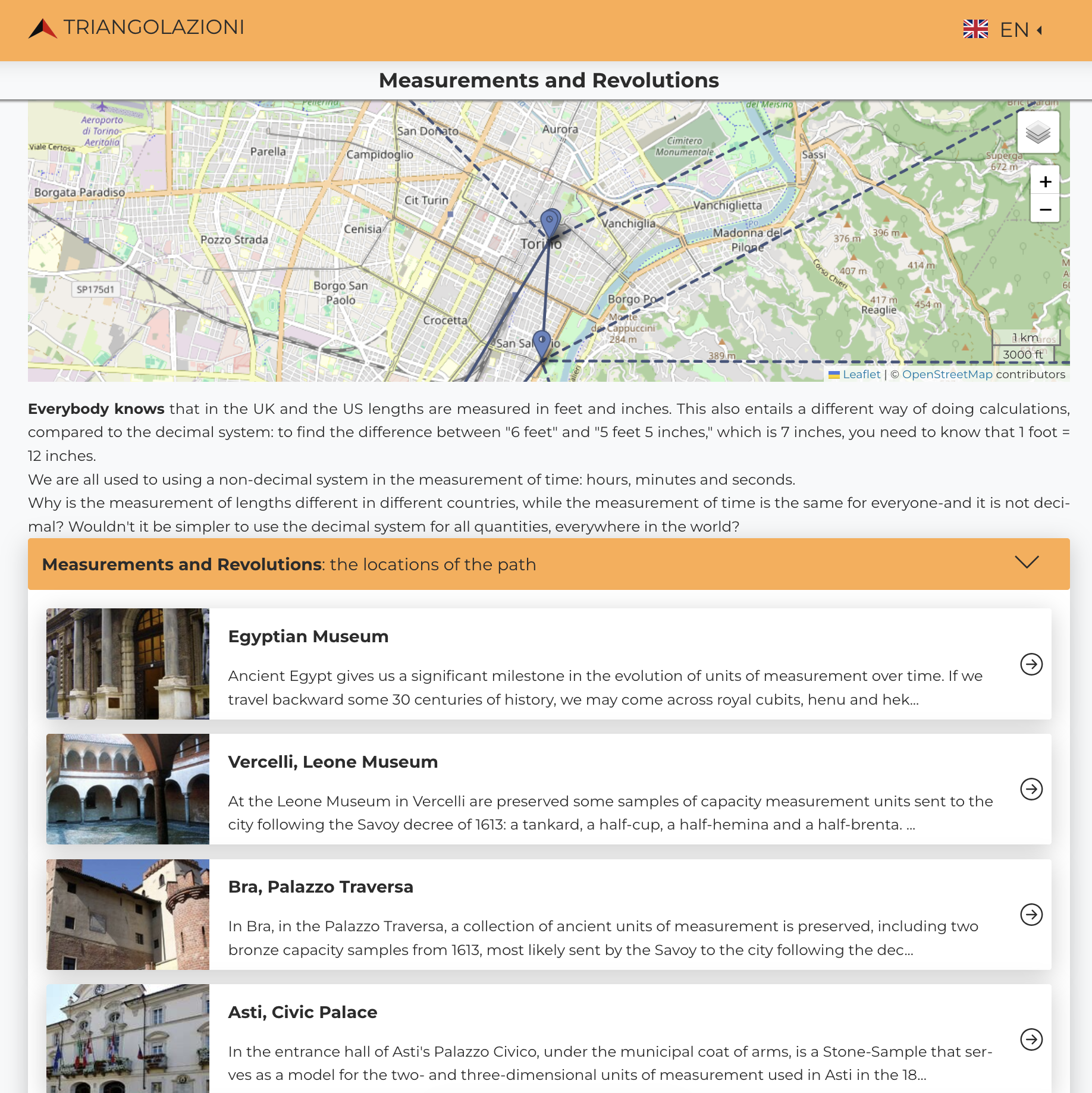}
         \caption{Presentation of a thematic path.}
         \Description[Measurements and Revolutions thematic path]{The upper portion of the page shows the thematic path on a geographical map, with the relevant Points of Interest, and describes the thematic path as a story. Below, a graphical component includes a link to each of these Points of Interest, with a small image, a short text, and a button to open the details page.}
         \label{fig:freeAndGenerated-a}
     \end{subfigure}
     \hfill
     \begin{subfigure}[b]{0.49\textwidth}
         \centering
         \includegraphics[width=\textwidth]{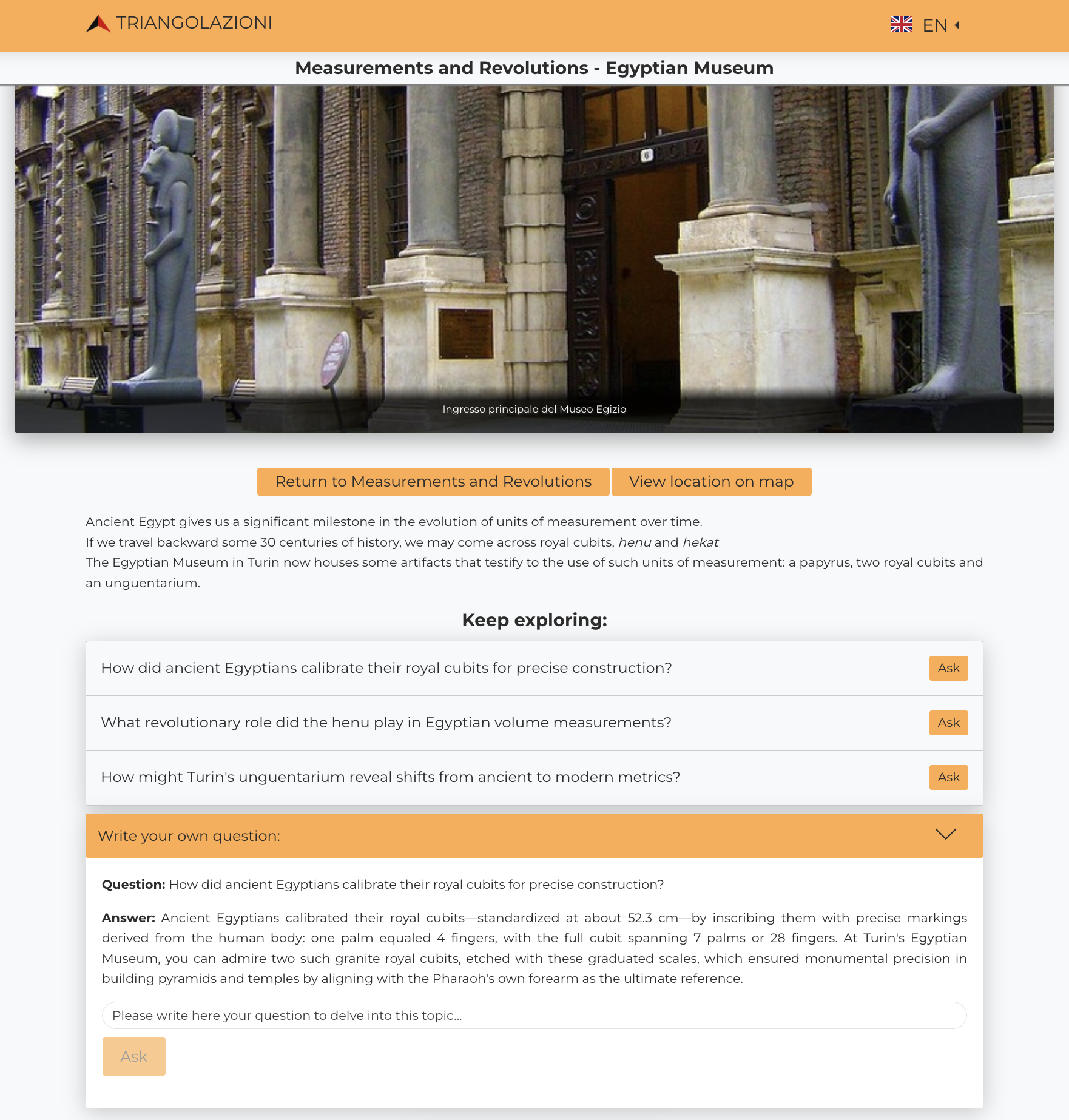}
         \caption{Suggested questions and chatbox's answer to the selected one.}
         \label{fig:freeAndGenerated-b}
     \end{subfigure}
        \caption{The user interface within the path Measurements and Revolutions (visualization for medium size screens).}
        \Description[Measurements and Revolutions - Egyptian Museum]{The upper portion of the page shows an image of the entrance of the museum. Below, there is a description of the museum and the graphical component showing the chatbot. This component includes three suggested questions and a text area to chat with the user.}
        \label{fig:freeAndGenerated}
\end{figure*}

% \section{Related work}
% \label{sec:related}

\section{The system}
\label{sec:model}
We build on the Triangolazioni CH guide \citep{Mauro-etal:22d}, which
%\citep{Mauro-etal:22d,Mauro-etal:22e}
presents thematic paths describing Points of Interest in Turin and its surroundings. Triangolazioni organizes thematic path as geolocalized narratives. It describes their places, historical personages, monuments, and historical or artistic objects. 
Its user interface is responsive and can be accessed from desktop and mobile devices. 

Our system extends Triangolazioni by integrating a chatbot powered by an LLM into its web pages.\footnote{The knowledge base of the original Triangolazioni guide is in Italian, and for the demonstration, we translated a part of it into English.} Figure \ref{fig:freeAndGenerated} shows two sample pages. The left one describes the ``Measurements and Revolutions'' thematic path, which focuses on measurement units (for time, space, etc.) and their evolution across time. The right one presents a Point of Interest (``Egyptian Museum'') in the context of the thematic path.

The chatbot works in the ``Keep exploring:'' area of the web page. An input field enables sending free-text questions (``Please write here your question to delve into this topic...''). The chatbot also proposes 3 personalized generated questions to suggest content exploration in greater depth. The user can write a question or submit the sample questions by clicking the ``Ask'' button, as in Figure \ref{fig:freeAndGenerated-b}. 

The system answers questions in the widget  ``Write your own question''. Below each answer, it shows another input field to let the user submit further questions, enabling a longer dialog.

The chatbot knows about the content of the current page and the previously browsed ones and answers the inquiries context-dependently. Therefore, both the suggested questions and the generated responses are personalized to the user's browsing history in the mobile guide.

Our system tracks the information that the user receives when interacting with it. By sharing this context with the chatbot, it instructs the underlying LLM to generate personalized questions and answers. 
For example, Figure \ref{fig:freeAndGenerated-b} shows the questions suggested in an interaction where the user only explored the ``Measurements and Revolutions'' path: ``How did ancient Egyptians calibrate their royal cubits for precise construction?'', ``What revolutionary role did the henu play in Egyptian volume measurements?'', and ``How might Turin's unguentarium reveal shifts from ancient to modern metrics?''.
In a different interaction, the user first explored some parts of the thematic path ``Measuring the Earth''
%, and the ``Palazzo Campana and Peano Library'' Point of Interest. In this interaction, the focus was 
focusing on mathematical aspects of spatial measurement. In that case, when moving to ``Measurements and Revolutions'' and inspecting the ``Egyptian Museum'', the suggested questions were: ``What ancient Egyptian tools in the museum helped standardize grain measurements?'', ``How did the royal cubit influence revolutionary changes in metric systems?'', ``Which papyrus artifact reveals Egypt's role in early revolutionary metrology?''.
These questions have a broader scope, referring to the standardization of measures over time.
%The answers depend on the interaction context, as well.

\section{Architecture}
\label{sec:prompt}

Our system is implemented using HTML and JavaScript and employs the Omeka-S semantic content management system\footnote{\url{https://omeka.org/s/}}) to store and manage curated data. The chatbot is powered by the \texttt{sonar} model provided by Perplexity\footnote{\url{https://www.perplexity.ai/api-platform/resources/meet-new-sonar}}
, which is built on top of Llama 3.3 70B. We used a temperature of 0 to ensure reproducibility.
It is integrated through implicit prompts shown in Table~\ref{tab:prompts}.

% Based on preliminary experiments, we configured the model with the following parameters: a temperature of 0 to ensure reproducibility; 
%\textit{top\_k} set to 5, corresponding to the number of candidate tokens retained during generation; and a presence penalty of 1. The latter parameter, whose range is $[-2.0, 2.0]$, encourages lexical novelty in the responses generated. A value of 1 introduces mild variability without affecting factual accuracy adversely.

\begin{table}[]
\caption{Prompt templates to interact with the chatbot.}
\Description[Prompt templates to interact with the chatbot]{There are five prompt templates. In each template, the variable part is enclosed in brackets.}
\label{tab:prompts}
%\resizebox{\textwidth}{!}{%
\resizebox{0.9\columnwidth}{!}{%
\begin{tabular}{ll}
\toprule
P1  & \begin{tabular}[c]{@{}l@{}}Hello, I am your digital tour guide for Turin. How can I assist you? \end{tabular} \\ 
\midrule
P2a & \begin{tabular}[c]{@{}l@{}}I am beginning this thematic path in the Turin area: [NAME OF\\ THE PATH]. Please tell me something about it.\end{tabular} \\
\midrule
P2b & \begin{tabular}[c]{@{}l@{}}I am looking at this place in the Turin area: [NAME OF THE\\ PLACE] in the path [NAME OF THE PATH]. Please tell me\\ something about it.\end{tabular} \\ 
\midrule
P3  & \begin{tabular}[c]{@{}l@{}}You are a tour guide. Please write 3 very short and engaging\\ questions about [PLACE NAME] in the thematic path [PATH \\NAME], based on our conversation so far, but whose answer is\\ not contained in it. Please, each line should contain only the\\ question, without quotation marks.\end{tabular} \\
\midrule
P4  & \begin{tabular}[c]{@{}l@{}}You are a tour guide of Turin. Please briefly answer the following\\ question, without repeating what we have said in this\\ conversation: [QUESTION]\end{tabular}
\\ \bottomrule
\end{tabular}%
}
\end{table}

To align the chatbot’s behavior with the user’s navigation of the website, the system maintains an interaction context that captures (i) the information presented to the user during web browsing and (ii) the responses previously generated by the chatbot. We represent this interaction context (hereafter denoted as \textit{CTX}) as a \textit{conversation} between the user and the system: the user's browsing activity is modeled as a dialog where the user asks for information and the guide responds by presenting the requested web pages. These pages include curated, static content and the dynamic content generated by the chatbot. CTX includes dialog turns representing the user's questions for the chatbot (LLM) and its responses.

Since the chatbot itself is stateless, an external software component is responsible for constructing and maintaining CTX. This component tracks both browsing actions and question-answer interactions. It supplies the full context to the chatbot at each invocation. The maximum size of CTX is constrained by the context window of the underlying LLM.

The integration between the web-based guide and the chatbot follows a loosely coupled architecture. A \textit{simulated user (SU)} and a \textit{simulated chatbot (SC)} act as proxies for the \textit{human user} and the \textit{LLM-powered chatbot}, respectively. SU and SC operate within the web browser and mediate the exchange of information between the website and the chatbot. They merge the content of visited web pages with the chatbot’s generated outputs into the shared conversation context CTX. 
The interaction flow proceeds as follows:
\begin{enumerate}
\item
When the user accesses the website for the first time, CTX is initialized with a welcome dialog act generated by the simulated chatbot SC and addressed to the simulated user SU (prompt P1 in Table~\ref{tab:prompts}).
\item
Each time the user navigates to a web page \textit{x}:
\begin{enumerate}
\item
The user interface provides SU with the URL and title of \textit{x}, which identifies a topic $T$ (e.g., ``Egyptian Museum'').
\item
To model the user’s intent, SU invokes SC using a prompt expressing interest in learning more about $T$. Prompt P2a/2b is used for pages describing thematic paths/Points of Interest.
\item
SC retrieves the content of page \textit{x} from its URL and appends it to CTX. This step unifies web navigation and question–answering by treating the page content as if it were provided by the LLM in response to SU’s request. As a result, CTX accurately reflects the information received by the user.
\item
To enable the generation of contextual question suggestions, SU asks SC to produce three candidate questions based on the current CTX (prompt P3).
\item
SC forwards the prompt together with CTX to the LLM through its API. In turn,
SC collects the generated questions and returns them to SU.
Finally, SU forwards the questions to the user interface, where they are displayed to the user.
\end{enumerate}
\end{enumerate}
For brevity, we omit the description of interactions triggered by explicit user questions and refer the reader to \citep{Geninatti-etal:25} for further details.

\section{Conclusions}
In a user study \citep{Geninatti-etal:25}, we found that our system has higher usability than the original Triangolazioni app. Moreover, its context-dependent question suggestions enhance user experience compared to free-text interaction with the LLM, especially for users with low levels in the Curiosity and Exploration Inventory-II \citep{Kashdan:09}.

%\textcolor{brown}{It can be noticed that, currently, LLMs cannot guarantee the quality and accountability of information because they can extract data from unreliable sources \citep{Mich-Garigliano:23}. However, recent studies show that their factuality tends to increase with their size \citep{Tam-etal:23}. This mitigates the problem in a question-answering task, as the one we aim to solve.}

This research extends the work of the Triangolazioni Project (\url{www.triangolazioni.unito.it/}), and has been funded by our University (Grant for Internationalization).
Code and data are available at \url{https://anonymous.4open.science/r/tell-me-more-A4E5/}.

\bibliographystyle{ACM-Reference-Format}
%\bibliography{plan-rec,mybib,httpbib} 

\begin{comment}
\newpage
\textbf{REQUIREMENTS FOR THE DEMONSTRATION}
%\textcolor{red}{On an extra page (not to be published), submissions should include a specification of the technical requirements for demonstrating the system at UMAP 2026.}

This demonstration will require a single table to place a laptop running the web application. Additionally, we will need a stable Wi-Fi Internet connection to access the application and interact with it, and an electric power connection.

If possible, an external monitor or screen would be beneficial to enhance visibility during the demonstration.
\end{comment}

\end{document}